\documentclass[12pt]{article}

\usepackage{amsmath,amssymb}
\numberwithin{equation}{section}
\usepackage{xfrac}
\usepackage{eqnarray} 
\usepackage{cancel}
\usepackage{algorithm}
\usepackage{algpseudocode}
\usepackage{bm}
\usepackage{syllogism}

\usepackage[T1]{fontenc}
\usepackage{mathpazo,mathabx}

\usepackage[normalem]{ulem}
\usepackage{sectsty}
\sectionfont{\fontsize{12}{15}\selectfont}
\subsectionfont{\normalfont\fontsize{12}{15}\selectfont}
\subsubsectionfont{\itshape\normalfont\fontsize{12}{15}\selectfont}

\usepackage{lineno}

\usepackage[symbol]{footmisc}

\usepackage{relsize}

\usepackage{accents}
\newlength{\dhatheight}

\newcommand*\patchAmsMathEnvironmentForLineno[1]{%
	\expandafter\let\csname old#1\expandafter\endcsname\csname #1\endcsname
	\expandafter\let\csname oldend#1\expandafter\endcsname\csname end#1\endcsname
	\renewenvironment{#1}%
	{\linenomath\csname old#1\endcsname}%
	{\csname oldend#1\endcsname\endlinenomath}%
}
\newcommand*\patchBothAmsMathEnvironmentsForLineno[1]{%
	\patchAmsMathEnvironmentForLineno{#1}%
	\patchAmsMathEnvironmentForLineno{#1*}%
}
\AtBeginDocument{%
	\patchBothAmsMathEnvironmentsForLineno{equation}%
	\patchBothAmsMathEnvironmentsForLineno{align}%
	\patchBothAmsMathEnvironmentsForLineno{flalign}%
	\patchBothAmsMathEnvironmentsForLineno{alignat}%
	\patchBothAmsMathEnvironmentsForLineno{gather}%
	\patchBothAmsMathEnvironmentsForLineno{multline}%
}

\usepackage{color}
\usepackage{wrapfig}
\usepackage{indentfirst}
\usepackage{setspace}
\usepackage[numbers,super,sort&compress]{natbib}
\usepackage{booktabs}
\usepackage{rotating}
\usepackage[margin = .85in]{geometry}
\usepackage{fancyhdr}
\usepackage{tabularx}
\usepackage{pdfpages}
\usepackage{longtable}
\usepackage{multirow}
\usepackage{multicol}
\usepackage{float}
\usepackage{lscape}
\usepackage{graphicx}
\usepackage{mdwlist}
\usepackage{url}
\usepackage{comment}
\usepackage{amsthm}

\usepackage{caption}
\usepackage{subcaption}

\usepackage[compact]{titlesec}

\makeatletter
\renewcommand\@biblabel[1]{#1.}
\makeatother

\usepackage[colorlinks=true, urlcolor=blue, linkcolor=black, citecolor=black]{hyperref}

\graphicspath{{../figures/}}

\newcommand{\repolink}{\href{https://github.com/ainaimi/LIE-primer}{GitHub repository}}

\begin{document}
\thispagestyle{empty}
\baselineskip=28pt

\noindent

\vskip 0.2cm {{\noindent \huge A Law of Iterated Expectation Primer for Causal Inference}}

\baselineskip=12pt

\vskip .25cm
\noindent Ashley I. Naimi, PhD$^{1,2 *}$\\[.1em]
\noindent Razieh Nabi, PhD$^{3}$\\[.1em]
\noindent Lindsay J. Collin, PhD$^{1}$\\[.1em]
\noindent Zhaohua Zeng, MSPH$^{1}$\\[.1em]
\noindent Paul N. Zivich, PhD$^{4}$\\[.1em]
\noindent Stephen R. Cole, PhD$^{4}$	\\[.1em]

\vskip .25cm
\noindent $^1$ Department of Epidemiology, Emory University.\\[.1em]
\noindent $^2$ Department of Data and Decision Sciences, Emory University.\\[.1em]
\noindent $^3$ Department of Biostatistics and Bioinformatics, Emory University.\\[.1em]
\noindent $^4$ Department of Epidemiology, UNC Chapel Hill.\\[.1em]

\vskip .25cm
\noindent \hskip -.2cm
\begin{tabular}{ll}
*Correspondence: & Department of Epidemiology \\[-.1cm]
& Rollins School of Public Health \\[-.1cm]
& Emory University \\[-.1cm]
& 1518 Clifton Road \\[-.1cm]
& Atlanta, GA 30322\\[-.1cm]
& \href{mailto:ashley.naimi@emory.edu}{ashley.naimi@emory.edu}
\end{tabular}

\begin{small}

\medskip
\noindent {\bf Publication status:} Accepted for publication in the \emph{American Journal of Epidemiology} (September 30, 2026).

\medskip
\noindent {\bf Conflicts:} The authors have no conflicts to disclose.

\medskip
\noindent {\bf Acknowledgements:} We thank Dr Edward Kennedy at CMU for sharing his causal inference course notes, on which some of this content was based. We thank Mr Zac Anderson-Cutting at Murdoch Children's Research Institute for identifying calculation errors in an initial version of this manuscript.

\medskip
\noindent {\bf Funding:} This work was supported by National Institutes of Health under award numbers K01AI177102 (PNZ), R01AI157758 (PNZ, SRC), and R01HL174652 (AIN). The content is solely the responsibility of the authors and does not necessarily represent the official views of the National Institutes of Health.

\medskip
\noindent {\bf AI Disclosure Statement:} Large language model tools (Claude Opus 4.8, Fable 5; Anthropic, San Francisco, CA) were used to check internal consistency, revise and organize the statistical code in the accompanying repository, and to  improve clarity of writing. The authors have verified the manuscript and code, and take full responsibility for its content.

\medskip
\noindent {\bf Data Availability Statement:} The data for Example 1 are reported in full in Table 1 of the manuscript. The NHEFS data used in Example 2 are publicly available with the book by Hern\'{a}n and Robins. The simulated data used in Example 3, the script that generates them (including the random seed), and R code reproducing all numerical results are available at \url{https://github.com/ainaimi/LIE-primer}.

\medskip

\end{small}

%
%
\newpage
\thispagestyle{empty}
\begin{center}
{\Large{\bf Abstract}} 
\end{center}
\baselineskip=12pt
The g-formula is a foundational tool for identifying causal effects in observational data. This tool is based  on the law of iterated expectation, a key mathematical identity in statistics. However, the notation with which the law of iterated expectation and the g-formula is expressed can be opaque to those with little background in statistics. We provide a primer introducing the law of iterated expectation, the integration notation used to express it, and its role for causal effect identification via the g-formula. Combined with the assumptions of causal consistency, positivity, and conditional exchangeability, the law of iterated expectation yields a causal standardization formula (the g-formula) in two nonparametrically equivalent forms: a non-iterative conditional expectation (NICE) form involving a single weighted average of conditional outcome means, and an iterative conditional expectation (ICE) form involving nested expectations. We illustrate both forms using three progressively complex numerical examples: a time-fixed example with a single binary confounder, a time-fixed example with discrete and continuous confounders, and a time-varying example with two timepoints. We provide clarity on what the law of iterated expectation is, how it is related to the g-formula, and how to gain intuition of its mathematical formulations in actual data examples that can be generalized to a range of settings.
\noindent  

%
%
%
\baselineskip=12pt
\par\vfill\noindent

\par\medskip\noindent
\newpage
\doublespacing
\setcounter{page}{1}

\section{Introduction}

Causal inference requires translating unobservable potential outcomes into functions of observed data, a step known as identification. This translation occurs as a mathematical exercise, and articulates some of the conditions needed to use data to estimate causal estimands defined in terms of potential outcomes.\cite{aronowNonparametricIdentificationNot2025} One mathematical tool central to this translation is the law of iterated expectation, an identity relating marginal and conditional means that appears throughout the causal inference literature. A firm grasp of this identity, and of the two equivalent forms in which it is typically expressed, makes the mechanics of widely used causal estimators, such as parametric g-formula, or g-computation, immediately transparent.

The g-formula was originally introduced by Robins in 1986\cite{Robins1986} as (i) a mathematical tool to identify causal effects using data, and (ii) a statistical ``plug in'' tool to estimate causal effects. With time, these distinct uses came to be characterized by different names: the ``g-formula'' for identification, and the ``g-computation algorithm'' for estimation. Another term used is the ``parametric g-formula'' or ``parametric g-computation'', which emphasizes the use of parametric models when needed to generate estimates with the procedure. Here, we maintain a distinction between identification (g-formula) and estimation (g-computation) throughout.

This primer seeks to build understanding from an intuitive standpoint, using concrete numerical examples to connect mathematics to practice. We explain connections between integration notation, the law of iterated expectation, and the g-formula for those with little background in mathematical statistics. To provide intuition, we rely on three increasingly complex applied examples.

\section{The Law of Iterated Expectation}

Throughout, uppercase letters such as $X$ and $Y$ denote random variables, while lowercase letters such as $x$ and $y$ denote particular realized values of those variables. The marginal expectation $E(Y)$ is the average value of $Y$ across the entire population. In contrast, the conditional expectation $E(Y \mid X = x)$ is the average value of $Y$ among individuals with the specific value $X = x$. When $Y$ is discrete, this conditional expectation is defined as $E(Y \mid X = x) = \sum_y y \cdot P(Y = y \mid X = x)$; when $Y$ is continuous with conditional probability density function $f_{Y \mid X}$, it is defined as $E(Y \mid X = x) = \int y\, f_{Y \mid X}(y \mid x)\,dy$. Viewed as a function of $x$, the quantity $E(Y \mid X = x)$ maps each possible value of $X$ to a corresponding conditional mean of $Y$. Evaluating this function at the random variable $X$ yields $E(Y \mid X)$, which is itself a random variable.

The law of iterated expectation (also known as the law of total expectation, the tower rule, or, for a binary $Y$ variable, the law of total probability) provides a formal connection between these quantities. The law states that, for any two random variables $Y$ and $X$ (with $E|Y| < \infty$):\cite{Wasserman2004}$^{\text{(p55)}}$
\begin{equation}
E(Y) = E[E(Y\mid X)]
\label{eqn:lie}
\end{equation}

\noindent In words, the marginal expectation of $Y$ can be recovered by first computing the conditional mean of $Y$ within strata of $X$, and then averaging the resulting random variable $E(Y \mid X)$ over the distribution of $X$. We refer to the right-hand side of equation \ref{eqn:lie} as the \emph{iterative} form, since it emphasizes the conditional expectation $E(Y \mid X)$ as a random variable whose expectation is then taken in turn.

The outer expectation in equation \ref{eqn:lie} is an average over the distribution of $X$ which, based on the ``law of the lazy statistician'',\cite{Wasserman2004}$^{\text{(p48)}}$ can be written as an average explicitly based on the definition of an expectation, giving the \emph{non-iterative} form:
\begin{equation}
E[E(Y\mid X)] = \int E(Y \mid X = x)\,d\mathbb{P}(x)
\label{eqn:lotus}
\end{equation}

\noindent where $\mathbb{P}$ denotes the probability distribution governing $X$. A first point of confusion is often the meaning of integrating ``with respect to a probability distribution $\mathbb{P}$ governing $X$.'' Informally, $\mathbb{P}(x)$ denotes the probability distribution induced by $X$, or the probability distribution that determines how the random variable $X$ behaves (e.g., as a categorical or continuous random variable). Integrating with respect to $\mathbb{P}(x)$ therefore corresponds to averaging over the distribution of $X$ (a more formal measure-theoretic explanation is provided in the Appendix).

A second point of confusion is that integrals such as $\int E(Y\mid X=x)\,d\mathbb{P}(x)$ can be written with different notation, but to the same effect (see Appendix). However, concretely the integral $\int E(Y\mid X=x)\,d\mathbb{P}(x)$ (or some of those in the Appendix) can be used as a single notational tool to convey that the average $E(Y \mid X = x)$ is weighted by the distribution of $X$. When $X$ is a categorical variable with $k$ categories, this integral collapses to a weighted sum:
\begin{align*}
	\int E(Y\mid X=x)\,d\mathbb{P}(x) \;=\; E(Y\mid X=1)\cdot & P(X=1)  \;+\; \\ E(Y\mid& X=2) \cdot P(X=2)  \;+\; \\ & \cdots \;+\; E(Y\mid X=k)\cdot P(X= k).
\end{align*}

\noindent When $X$ is continuous with probability density function $f_X$, the non-iterative form can be written as:
\begin{equation}
\int E(Y\mid X=x)\,d\mathbb{P}(x) \;=\; \int_{-\infty}^{\infty} E(Y\mid X=x)\,f_X(x)\,dx.
\end{equation}

\noindent The density $f_X(x)$ plays the role of the weights, and, instead of a discrete sum, a continuous integral is implied. However, the intuition is the same as in the binary case: we are averaging the conditional mean $E(Y\mid X=x)$ over all values of $x$, weighted by how probable each value is under the distribution of $X$.

In many technical papers, integrals written with notation such as $d\mathbb{P}(x)$ are referred to as ``Lebesgue,'' or ``Lebesgue--Stieltjes'' integrals (see Appendix). In this primer, the integrals we write can be evaluated as a weighted sum for discrete variables, or a density-weighted integral for continuous ones. Notation such as $\int E(Y \mid X = x)\,d\mathbb{P}(x)$ accommodates both scenarios, and appears throughout the causal inference literature. The Appendix provides a self-contained overview of the different integration concepts and the measure-theoretic ideas that unify them; readers seeking deeper treatments can consult standard texts.\cite{Vaart2000, Pollard2002} For the examples considered in this primer (as in most applied work), the distinctions between these and other integrals do not affect the resulting numerical quantities, but can matter for theoretical work involving asymptotic convergence, consistency, and efficiency theory.\cite{Vaart2000}

\section{Example 1: Time Fixed Example}

\begin{wrapfigure}{r}{0.45\textwidth}
\centering
  \begin{minipage}{0.40\textwidth}
    \centering
    \includegraphics[scale = .4]{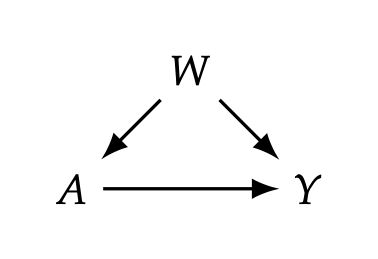}
    \caption{A directed acyclic graph (DAG) illustrating the causal structure among treatment $A$, outcome $Y$, and baseline covariate $W$. An arrow from one node to another indicates a direct causal effect. The path $A \leftarrow W \to Y$ represents confounding of the effect of $A$ on $Y$ by $W$. Alternative text: a graph with three nodes labeled $W$, $A$, and $Y$, with arrows pointing from $W$ to $A$, from $W$ to $Y$, and from $A$ to $Y$.}
    \label{fig:dag_a}
  \end{minipage}%
\end{wrapfigure}

We begin with an applied example provided by Sato and Matsuyama,\cite{Sato2003} with data on the effect of tamoxifen use on breast cancer recurrence among women who underwent surgery for breast cancer. These observational data include 4,901 women with breast cancer who used tamoxifen (A=1) or did not (A=0), with an outcome measure of breast cancer recurrence (Y=1). Women were also classified as having positive lymph node metastasis (W=1) or not (W=0) at the time of surgery before tamoxifen use. The causal diagram is represented by the directed acyclic graph (DAG) depicted in Figure \ref{fig:dag_a}.

 Causal effects are often expressed as contrasts, such as the causal risk difference: $\psi = E(Y^{a=1} - Y^{a=0})$, where $Y^a$ is the potential outcome that would be observed if $A$ were set to $a$. However, potential outcomes are not directly observable. Identification assumptions allow us to mathematically equate $E(Y^{a})$ with some function of our observed data, which means we can use averages of observed outcomes to quantify averages of potential outcomes.\cite{Naimi2023a,Hernan2025}$^{\text{(chpt 2)}}$

For a dataset generated from the DAG in Figure \ref{fig:dag_a}, one sufficient set of identification assumptions includes causal consistency, positivity, and conditional exchangeability.
In practice, these assumptions may additionally require appropriate adjustment for selection processes or measurement error. 
For data generated from Figure \ref{fig:dag_a} the latter assumption requires adjusting for variables $W$. In this example, $W$ is a single binary variable (i.e., $W \in \{0,1\}$). Together, these assumptions allow us to do the following: 
\begin{align}
 E(Y^a) & = \sum_w E(Y^a \mid W = w)\cdot P(W = w)  \\
        & = \sum_w E(Y^a \mid A = a, W = w)\cdot P(W = w) \\
        & = \sum_w E(Y \mid A = a, W = w)\cdot P(W = w) \label{eqn:last}
\end{align}

\noindent where the first equality is given by the law of iterated expectation, the second by conditional exchangeability,\cite{Naimi2023a} the third by causal consistency (including treatment variation irrelevance and no interference).\cite{Vanderweele2009a} Positivity ensures that the conditional expectations in the resulting observed-data functional are identifiable from the observed data. Under these identification assumptions, the last equation \ref{eqn:last} only involves observed variables, but is equal to $E(Y^a)$, allowing us to estimate this causal object with our data. 

\medskip
\noindent\fbox{%
  \begin{minipage}{\dimexpr\linewidth-2\fboxsep-2\fboxrule\relax}
    \smallskip
    \textbf{Box 1: The G-Formula versus the Law of Iterated Expectation}\\[4pt]
The law of iterated expectation is a statistical statement about the relationships between random variables. For example, if we refer to the DAG in Figure \ref{fig:dag_a}, we can write:
\begin{equation}
E(W) = E[E(W \mid Y)] = \sum_y E(W \mid Y = y)\cdot P(Y = y)
\end{equation}
 which would be valid as a statistical relation. However, the g-formula is a statement about causality. It starts by stipulating a probability model for the joint distribution of the data and factoring this distribution according to the \emph{causal} ordering of the variables encoded in the DAG. After converting the factored joint density to a sequence of expectations, and intervening to yield a degenerate distribution for the exposure (which allows us to drop any terms for the propensity score), we then get the more familiar ``law of iterated expectation'' version of the g-formula:
 \begin{equation}
 E(Y^a) = \int E(Y \mid A = a, W = w)\,d\mathbb{P}(w)
 \end{equation}
 Thus, the law of iterated expectation is a mathematical identity that holds for any probability distribution. The g-formula is obtained by combining the law of iterated expectation with causal identification assumptions that permit replacement of counterfactual quantities by observed data quantities.
    \smallskip
  \end{minipage}%
}
\medskip

\subsection{Non-Iterative Conditional Expectation}

Note that the right-hand side of equation \ref{eqn:last} uses the law of iterated expectation (a weighted average of conditional means) with its causal interpretation resting on the identification assumptions. Just as the law of iterated expectation can be written in a non-iterative and an iterative form (equations \ref{eqn:lie} and \ref{eqn:lotus}), the identified functional in equation \ref{eqn:last} can be expressed in two equivalent ways, referred to as the ``non-iterative conditional expectation'' (NICE) and ``iterative conditional expectation'' (ICE) versions of the g-formula:\cite{Wen2021}
\begin{equation}
E(Y^a) = \underbrace{\sum_w E(Y \mid A = a, W = w)\cdot P(W = w)}_{\text{NICE}} = \underbrace{E[E(Y \mid A = a, W)]}_{\text{ICE}}
\label{eqn:niceice}
\end{equation}

\noindent We illustrate each in turn, beginning with the NICE form. With a single binary confounder, the NICE g-formula (equivalently, $\int E(Y \mid A = a, W = w)\,d\mathbb{P}(w)$ in integral notation) is a simple sum:

\begin{align*}
E(Y^{a}) & = \sum_w E(Y \mid A = a, W = w)\cdot P(W = w)\\
         & = E(Y \mid A = a, W = 1)\cdot P(W = 1) +  E(Y \mid A = a, W = 0)\cdot P(W=0)
\end{align*}

We can use the data in Table \ref{tab:tamoxifen_data} to construct a ``plug-in'' estimator of $E(Y^{a})$, assuming the identification assumptions hold. From Table \ref{tab:tamoxifen_data}, an estimate of $E(Y^{a=1})$ can be computed by first estimating the mean of breast cancer recurrence among women with $A = 1$ and $W = 1$, the mean of breast cancer recurrence among women with $A = 1$ and $W = 0$, as well as the overall proportion of women with $W = 1$ and $W = 0$. 

\begin{table}[h]
\centering
\begin{tabular}{ccrr}
\toprule
$W$ & $A$ & $Y$ & $N$ \\
\midrule
0 & 0 & 0 & 1{,}421 \\
0 & 0 & 1 &    171 \\
0 & 1 & 0 & 1{,}238 \\
0 & 1 & 1 &     96 \\
1 & 0 & 0 &    507 \\
1 & 0 & 1 &    253 \\
1 & 1 & 0 &    847 \\
1 & 1 & 1 &    368 \\
\bottomrule
\end{tabular}
\caption{Data from Sato and Matsuyama (2003) on the effect of tamoxifen use on breast cancer recurrence among women who underwent surgery for breast cancer. These observational data include 4901 women with breast cancer who used tamoxifen (A=1) or did not (A=0), with an outcome measure of breast cancer recurrence (Y=1) or not (Y=0). Women were also classified as having positive lymph node metastasis (W=1) or not (W=0) at the time of surgery before tamoxifen use.}
\label{tab:tamoxifen_data}
\end{table}

\newpage

Using Table \ref{tab:tamoxifen_data}, we get:

\begin{itemize}
	\item[] $\hat{E}(Y \mid A = 1, W = 1)$: $\frac{368}{(847+368)} = 0.303$
	\item[] $\hat{E}(Y \mid A = 1, W = 0)$: $\frac{96}{(96+1{,}238)} = 0.072$
	\item[] $\hat{P}(W = 1)$: $\frac{(507+253+847+368)}{(1{,}421+171+1{,}238+96+507+253+847+368)} = \frac{1{,}975}{4{,}901} = 0.403$
	\item[] $\hat{P}(W = 0)$: $1 - 0.403 = 0.597 $
\end{itemize}

\noindent and we can ``plug'' these numbers into the expression for $E(Y^{a = 1})$ to get:

\begin{equation}
0.303 \times 0.403 + 0.072 \times 0.597 \approx 0.165
\end{equation}

We repeat the process for $E(Y^{a = 0})$, which yields:

\begin{itemize}
    \item[] $\hat{E}(Y \mid A = 0, W = 1)$: $\frac{253}{(507+253)} = 0.333$
    \item[] $\hat{E}(Y \mid A = 0, W = 0)$: $\frac{171}{(171+1{,}421)} = 0.107$
\end{itemize}

\noindent which, when plugged into the expression for $E(Y^{a = 0})$:

\begin{equation}
0.333 \times 0.403 + 0.107 \times 0.597 \approx 0.198
\end{equation}

Giving:

\begin{equation}
\hat{\psi} = 0.165 - 0.198 \approx -0.033
\end{equation}

\subsection{Iterative Conditional Expectation}

The ICE version of the g-formula in equation \ref{eqn:niceice} re-expresses the same identified quantity as a set of nested expectations:

\begin{equation}
E(Y^a) = E[E(Y \mid A = a, W)].
\end{equation}

The first (inner) expectation represents predicted values of $Y$ with $A$ fixed at $a$ for those with the observed values of $W$. The second (outer) expectation represents the average of these predicted values of $Y$ over the observed distribution of $W$. Using the Table \ref{tab:tamoxifen_data} data, we have already computed estimates of $E(Y \mid A, W)$ for each stratum of $A$ and $W$:

\begin{itemize}
  \item[] $\hat{E}(Y \mid A = 1, W = 1)$: $\frac{368}{(847+368)} = 0.303$
  \item[] $\hat{E}(Y \mid A = 1, W = 0)$: $\frac{96}{(96+1{,}238)} = 0.072$
  \item[] $\hat{E}(Y \mid A = 0, W = 1)$: $\frac{253}{(507+253)} = 0.333$
  \item[] $\hat{E}(Y \mid A = 0, W = 0)$: $\frac{171}{(171+1{,}421)} = 0.107$
\end{itemize}

A natural approach is to construct individual-level predicted values $\widetilde{Y}_a$ for each person at each treatment level, assigning each the conditional mean $\hat{E}(Y\mid A = a,W)$ at their observed covariate values:

\begin{table}[h]
\centering
\begin{tabular}{cccccc}
\toprule
ID &  $W$ & $A$ & $Y$ & $\widetilde{Y}_1$ & $\widetilde{Y}_0$ \\
\midrule
1 & 1 & 1 & 0 &  0.303 & 0.333 \\
2 & 0 & 1 & 1 &  0.072 & 0.107 \\
3 & 1 & 0 & 0 &  0.303 & 0.333 \\
4 & 1 & 0 & 1 &  0.303 & 0.333 \\
5 & 0 & 1 & 0 &  0.072 & 0.107 \\
6 & 1 & 0 & 1 &  0.303 & 0.333 \\
7 & 0 & 0 & 0 &  0.072 & 0.107 \\
8 & 0 & 1 & 1 &  0.072 & 0.107 \\
\vdots & \vdots &  \vdots & \vdots &  \vdots &  \vdots \\
\bottomrule
\end{tabular}
\caption{Extract of individual-level data from the Sato and Matsuyama (2003) dataset augmented with stratum-specific predicted outcome values $\widetilde{Y}_a$ for $a \in \{0, 1\}$. For each individual, $\widetilde{Y}_a$ contains the conditional mean $\hat{E}(Y \mid A = a, W)$ evaluated at the treatment level $a$ and that individual's covariate values. These predicted values are averaged in the outer expectation of the iterative conditional expectation g-formula.}
\label{tab:ice_illustration}
\end{table}

\newpage

Averaging $\widetilde{Y}_a$ computes the outer expectation in the ICE expression identifying $E(Y^a)$. For $a=1$:

\begin{equation}
    \hat{E}[\hat{E}(Y \mid A = 1, W)] = \frac{1}{N}\sum_{i = 1}^{N} \widetilde{Y}_{1i}
\label{eqn:iceouter}
\end{equation}
\noindent where $i = 1 \ldots N$ indexes \emph{all} individuals in the sample, regardless of their observed exposure. In Table \ref{tab:ice_illustration}, $\widetilde{Y}_1$ takes the value $0.303$ for each of the $1{,}975$ women with $W = 1$ (whether treated or not) and $0.072$ for each of the $2{,}926$ women with $W = 0$. Equation \ref{eqn:iceouter} therefore evaluates to:
\begin{equation*}
    \frac{0.303 \times 1{,}975 + 0.072 \times 2{,}926}{4{,}901} \approx 0.165
\end{equation*}

\noindent exactly as in the NICE implementation. The outer average is not taken within the exposed stratum, but rather by averaging the predicted outcomes $\widetilde{Y}_1$ and $\widetilde{Y}_0$ over all individuals in the sample. Repeating the full-sample average in equation \ref{eqn:iceouter} for $\widetilde{Y}_0$ yields $0.198$, and thus:
\begin{equation}
    \hat{\psi} = 0.165 - 0.198 \approx -0.033
\end{equation}

\noindent which matches the estimate obtained from the NICE implementation.

\section{Example 2: More Complex Time-Fixed Example}

Our next example still follows the DAG in Figure \ref{fig:dag_a}, but $W$ is a  multivariate vector with one continuous, two binary, and one categorical confounder. The example is based on the data from the NHANES Epidemiologic Follow Up Survey (NHEFS), available with the book by Hern\'{a}n and Robins.\cite{Hernan2025} Our goal is to estimate the effect of quitting smoking ($A = 1$) on weight change between 1971 and 1982 ($Y$). Our analytic sample includes 1,394 men and women who were smokers in 1971 and who had complete data on smoking cessation, weight change, and the four covariates used here; for pedagogical simplicity, we treat missing data as ignorable. The confounders are age in 1971 (continuous), family income in 1971 (recoded from the NHEFS 12-level income variable into three categories: low [codes 11--16], middle [17--19], and high [20--22]), race (binary), and sex (binary). This deliberately reduced covariate set is chosen for illustration, the resulting estimates should not be interpreted as substantive findings. Throughout this example, the outcome model is a linear regression of weight change on quit-smoking status, age (entered linearly), income, race, and sex, with no product terms. The first eight rows of the data are provided in Table \ref{tab:nhefs_data}.

\begin{table}[ht]
\centering
	\begin{tabular}{rrrrrrr}
  \hline
 ID & A & Y & age & income & race & sex \\ 
  \hline
  1 & 0.00 & -10.09 & 42.00 & Middle & 1.00 & 0.00 \\ 
  2 & 0.00 & 2.60 & 36.00 & Middle & 0.00 & 0.00  \\ 
  3 & 0.00 & 9.41 & 56.00 & Low    & 1.00 & 1.00 \\ 
  4 & 0.00 & 4.99 & 68.00 & Low    & 1.00 & 0.00 \\ 
  5 & 0.00 & 4.99 & 40.00 & Middle & 0.00 & 0.00 \\ 
  6 & 0.00 & 4.42 & 43.00 & Low    & 1.00 & 1.00 \\ 
  7 & 0.00 & -2.72 & 51.00 & Middle & 0.00 & 0.00 \\ 
  8 & 0.00 & 9.86 & 43.00 & Low    & 0.00 & 0.00 \\  
   \hline
 	\end{tabular}
	\caption{Extract of the data from the NHANES Epidemiologic Follow Up Survey on the effect of quitting smoking on weight change between 1971 and 1982. These observational data include 1394 men and women who were smokers at baseline, some of whom quit smoking over the course of the follow-up period ($A = 1$), with an outcome measure of the difference in weight (in kg) between 1982 and 1971 ($Y$). Example confounding variables include age (continuous), income (three categories), race (binary), and sex (binary).}
   \label{tab:nhefs_data}
\end{table}

\subsection{Non-Iterative Conditional Expectation}

In this example, because $W$ represents a vector of mixed type confounders, one cannot express the NICE g-formula as a simple sum of weighted outcome averages. Let $W_1$ denote age (continuous), $W_2$ race (binary), $W_3$ sex (binary), and $W_4$ income (three categories). Writing the NICE g-formula out more completely, $E(Y^a)$ is identified (under the causal assumptions) by:

\begin{align*}
\sum_{w_4=0}^{2}\sum_{w_3=0}^{1}\sum_{w_2=0}^{1}\Bigg(\int_{\mathcal{W}_1} & E(Y \mid A=a, W_1 = w_1, W_2 = w_2, W_3 = w_3, W_4 = w_4)\\
	   & \times f_{W_1 \mid W_2, W_3, W_4}(w_1 \mid w_2, w_3, w_4)\, dw_1 \Bigg)\times P(W_2=w_2, W_3=w_3, W_4=w_4)
\end{align*}

\noindent where $f_{W_1 \mid W_2, W_3, W_4}$ is the conditional density of age given the three discrete confounders, and $\mathcal{W}_1$ is the set of possible age values. This expression arises from factoring the joint distribution of the confounders as $f_{W_1 \mid W_2, W_3, W_4}(w_1 \mid w_2, w_3, w_4) \times P(W_2=w_2, W_3=w_3, W_4=w_4)$. Any factorization of this joint distribution would be valid, but isolating the continuous covariate in this way makes explicit which part of the weighted average requires integration over a density (age) and which parts reduce to simple sums (race, sex, and income).

In principle, evaluating this sum-integral combination in data could be done, but it would be laborious (relative to the iterated conditional expectation form). In code, this requires fitting an outcome model, predicting $Y$ under each exposure value, averaging predictions within strata of the discrete confounders, then computing a weighted sum across strata. 

For clarity, we provide R code to implement this in the \repolink{} associated with this article. Taking this example to completion for illustration, we obtain:
\begin{align*}
	\hat{E}(Y^{a = 1}) & = \int \hat{E}(Y \mid A = 1, W = w)\,d\hat{\mathbb{P}}(w) = 5.0 \\
	\hat{E}(Y^{a = 0}) & = \int \hat{E}(Y \mid A = 0, W = w)\,d\hat{\mathbb{P}}(w) = 1.9 \\
	\hat{\psi} & = \hat{E}(Y^{a = 1}) - \hat{E}(Y^{a = 0}) = 3.1 \\
\end{align*}

\subsection{Iterative Conditional Expectation}

We could implement this same procedure more simply using the iterative conditional expectation approach as well. For example, we can write:

\begin{equation}
E(Y^a) = E_W[E(Y \mid A = a, W_1, W_2, W_3, W_4)]
\end{equation}

\noindent where $E_W(\bullet)$ denotes taking the expectation over the entire joint distribution of $W_1, W_2, W_3, W_4$ (note that the $E_W()$ notation, where the $W$ is subscripted, is often left out of this type of equation). This iterative conditional expectation procedure is trivially easy to implement in code. For example, the procedure could be implemented as: 

\begin{itemize}
	\item[1.] Fitting a regression model for the mean of $Y$ conditional on $A$, and all $W$'s
	\item[2.] Generating predictions from this model for $Y$ under $A = 1$ and $A = 0$ for all individuals in the sample
	\item[3.] Averaging these predictions in the sample (i.e., over the joint distribution of $W$'s)
	\item[4.] Taking the difference of these averaged predictions
\end{itemize}

This procedure approximates the population integral by averaging over the empirical distribution of the observed covariates. R code to implement this procedure is available in the associated \repolink{}.

Implementing this in the NHEFS data with code, we obtain exactly the same results as in the NICE implementation:
\begin{align*}
\hat{E}(Y^{a = 1}) & = \hat{E}[\hat{E}(Y \mid A = 1, W)] = 5.0 \\
  \hat{E}(Y^{a = 0}) & = \hat{E}[\hat{E}(Y \mid A = 0, W)] = 1.9 \\
	\hat{\psi} & = \hat{E}(Y^{a = 1}) - \hat{E}(Y^{a = 0}) = 3.1
\end{align*}

Both approaches yield $\hat{\psi} = 3.1$ kg. In this time-fixed setting, the agreement is exact by construction: both implementations average the same outcome-model predictions over the same empirical covariate distribution (NICE by averaging within cells of the discrete confounders and then weighting by cell proportions, ICE by averaging the predictions directly over the whole sample). Because we did not need to fit any model for the confounder distribution (the empirical distribution was used in both cases), the two computations must return the same number. As we discuss below, this will not always hold in time-varying settings, specifically where the NICE implementation requires parametric models for the time-varying confounders.

\section{Example 3: Time Varying Example}

\begin{wrapfigure}{r}{0.55\textwidth}
\vspace{6pt}
\centering
  \begin{minipage}{0.51\textwidth}
    \centering
    \includegraphics[scale = .4]{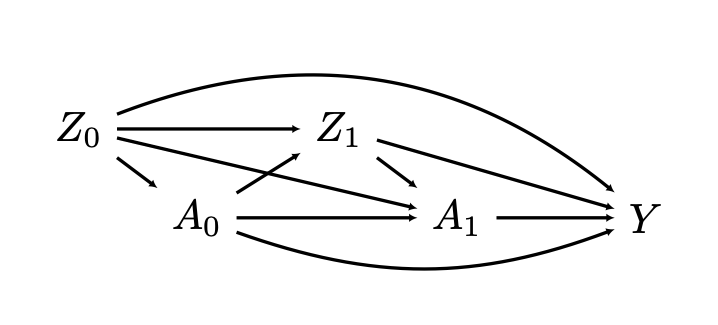}
    \caption{A DAG illustrating the causal structure for a time-varying confounding structure. The treatment and time-varying confounding variables $A$ and $Z$ are measured at two time-points $t \in \{0,1\}$. The outcome $Y$ is measured at the end of follow-up. This structure implies that: (a) the effect of $A_1$ on $Y$ is confounded by $Z_1$, $Z_0$, and $A_0$; and (b) the effect of $A_0$ on $Y$ is confounded by $Z_0$. Furthermore, part of the effect of $A_0$ on $Y$ is mediated by $Z_1$ and $A_1$. Because of this, one cannot simply fit a standard regression model to estimate the effect of setting $A_0, A_1$ to some specific values $a_0, a_1$. Instead, g methods can be used, such as the g-formula, longitudinal AIPW, or longitudinal TMLE, all of which invoke the law of iterated expectation in a sequential manner. Alternative text: a graph with five nodes labeled $Z_0$, $A_0$, $Z_1$, $A_1$, and $Y$, ordered in time, with arrows pointing from each variable to the variables that follow it.}
    \label{fig:dag_b}
  \end{minipage}%
\vspace{6pt}
\end{wrapfigure}

The law of iterated expectation becomes essential in time-varying settings where standard regression fails. We suppose a time-varying data generating structure as displayed in Figure \ref{fig:dag_b}. This Figure shows a time-varying treatment $A$ and a time-varying confounder affected by prior exposure $Z$, both measured at two time-points: $t \in \{0, 1\}$. The outcome $Y$ is measured at the end of follow-up. This structure implies that: (a) the effect of $A_1$ on $Y$ is confounded by $Z_1$, $Z_0$, and $A_0$; and (b) the effect of $A_0$ on $Y$ is confounded by $Z_0$. Furthermore, part of the effect of $A_0$ on $Y$ is mediated by $Z_1$ and $A_1$. Because of this, one cannot simply fit a standard regression model to estimate the effect of setting $A_0, A_1$ to some specific values $a_0, a_1$.\cite{Naimi2017a}

Suppose in this case we were interested in the average treatment effect of $A$ on $Y$, defined as:
\begin{align*}
	\psi & = E(Y^{a_0 = 1, a_1 = 1} - Y^{a_0 = 0, a_1 = 0})\\
	     &= E(Y^{\overline{a}_1 = 1} - Y^{\overline{a}_1 = 0})
\end{align*}
\noindent where $Y^{\overline{a}_1}$ is the potential outcome that would be observed if we set the history of the exposure up to time-point 1 (denoted $\overline{a}_1 = \{a_0, a_1\}$) to some specific values. 

For a dataset generated from a DAG in Figure \ref{fig:dag_b}, we can rely on an extension of the same set of identification assumptions including causal consistency, sequential positivity, and sequential conditional exchangeability. As in the time-fixed example, under these assumptions, we can do the following:
\begin{align}
E(Y^{\overline{a}_1}) & = \int \int E(Y \mid A_1 = a_1, A_0 = a_0, Z_1 = z_1, Z_0 = z_0)\,d\mathbb{P}(z_1\mid a_0, z_0)d\mathbb{P}(z_0) \label{eq:nicegform1}\\
					  & = E \bigg \{ E \bigg [ E \big (Y \mid A_1 = a_1, A_0 = a_0, Z_1, Z_0 \big ) \, \bigg| \, A_0 = a_0, Z_0 \bigg ] \bigg \} \label{eq:icegform1}
\end{align}
Here, equation \ref{eq:nicegform1} represents the conditional expectation of the outcome $Y$, weighted by the product of the conditional distribution of $Z_1$, and (because nothing affects $Z_0$ in the DAG) the marginal distribution of $Z_0$. Similarly, equation \ref{eq:icegform1} represents a sequence of nested predictions, subsequently averaged over all individuals in the sample. We explain each in turn. The data we use to do so were simulated ($n = 10^6$ individuals) from a data generating mechanism following the causal relations depicted in Figure \ref{fig:dag_b}, adapted from Naimi et al.\cite{Naimi2017a} Table \ref{tab:time_varying_data} summarizes the simulated data, reporting the mean of the continuous outcome $Y$ and the number of individuals within each of the 16 strata defined by the four binary time-varying variables (two exposures, $A_0$, $A_1$; and two time-varying confounders, $Z_0$, $Z_1$). We provide the full individual-level dataset as an accompanying .csv file, along with the generating script (including the random seed), in the associated \repolink{}.

\begin{table}[ht]
\centering
\begin{tabular}{rrrrrr}
  \hline
 $Z_0$ & $A_0$ & $Z_1$ & $A_1$ & $Y$ & $N$ \\ 
  \hline
    0 &   0 &   0 &   0 & 150.00 & 195921 \\
    0 &   0 &   0 &   1 & 152.00 & 72454 \\
    0 &   0 &   1 &   0 & 148.50 & 63466 \\
    0 &   0 &   1 &   1 & 150.49 & 34671 \\
    0 &   1 &   0 &   0 & 151.50 & 55600 \\
    0 &   1 &   0 &   1 & 153.50 & 30959 \\
    0 &   1 &   1 &   0 & 149.99 & 26039 \\
    0 &   1 &   1 &   1 & 152.00 & 21706 \\
    1 &   0 &   0 &   0 & 151.50 & 133412 \\
     1 &   0 &   0 &   1 & 153.50 & 73492 \\
     1 &   0 &   1 &   0 & 150.00 & 62708 \\
     1 &   0 &   1 &   1 & 152.00 & 51950 \\
     1 &   1 &   0 &   0 & 153.00 & 43432 \\
     1 &   1 &   0 &   1 & 155.00 & 35710 \\
     1 &   1 &   1 &   0 & 151.50 & 43843 \\
     1 &   1 &   1 &   1 & 153.51 & 54637 \\
   \hline
\end{tabular}
\caption{Tabled data demonstrating the mean of a continuous outcome variable $Y$ within all strata created by four time-varying variables, along with the number of individuals ($N$) in each stratum. These data comprise a single draw of $n = 10^6$ individuals generated from a simulation mechanism following the causal relations depicted in Figure \ref{fig:dag_b}; the generating script and random seed are provided in the accompanying repository, so that all quantities reported in this section can be reproduced exactly from the tabled (or individual-level) data.}
\label{tab:time_varying_data}
\end{table}

\subsection{NICE G-Formula with Two Timepoints}

The non-iterative conditional expectation g-formula with two timepoints, as expressed in the above equation \ref{eq:nicegform1}, includes the conditional expectation of the outcome $Y$ as a function of both $Z$ and $A$ variables, weighted by the distributions of $Z_1$ and $Z_0$. Because both confounders are binary, the integrals in equation \ref{eq:nicegform1} evaluate to the double sum:
\begin{align*}
E(Y^{\overline{a}_1}) & = \sum_{z_0 = 0}^1 \sum_{z_1=0}^1 E(Y \mid Z_0 = z_0, A_0 = a_0, Z_1 = z_1, A_1 = a_1)P(Z_1 = z_1 \mid Z_0 = z_0, A_0 = a_0)P(Z_0 = z_0)
\end{align*}
In this equation, the law of iterated expectation can be implemented by taking the average of $Y$ conditional on $Z_0 = z_0$, $A_0 = a_0$, $Z_1 = z_1$, and $A_1 = a_1$, weighting these conditional means by the probabilities of $Z_1 = z_1$ given $Z_0 = z_0$ and $A_0 = a_0$, and then weighting by the probabilities of $Z_0 = z_0$.

The conditional means $\hat{E}(Y \mid Z_0, A_0, Z_1, A_1)$ are read directly from Table~\ref{tab:time_varying_data}. Two additional quantities are required: the marginal probability $\hat{P}(Z_0)$ and the conditional probabilities $\hat{P}(Z_1 \mid Z_0, A_0)$, which can be computed using the $N$s in the tabled data. For example, $\hat{P}(Z_1 = 1 \mid Z_0 = 0, A_0 = 0)$ is the proportion of individuals with $Z_1 = 1$ among those with $Z_0 = 0$ and $A_0 = 0$: $(63{,}466+34{,}671)/(195{,}921+72{,}454+63{,}466+34{,}671) = 0.268$. Proceeding this way for all required quantities:

\begin{itemize}
  \item $\hat{P}(Z_0 = 1) = 0.499$
  \item $\hat{P}(Z_1 = 1 \mid Z_0 = 0, A_0 = 0) = 0.268$;\quad $\hat{P}(Z_1 = 1 \mid Z_0 = 0, A_0 = 1) = 0.355$
  \item $\hat{P}(Z_1 = 1 \mid Z_0 = 1, A_0 = 0) = 0.357$;\quad $\hat{P}(Z_1 = 1 \mid Z_0 = 1, A_0 = 1) = 0.554$
\end{itemize}

Plugging these into the NICE g-formula for $\overline{a}_1 = (1,1)$, each of the four $(z_0, z_1)$ combinations contributes one term, with weights $\hat{P}(Z_1 = z_1 \mid Z_0 = z_0, A_0 = 1)$ [e.g., $0.645 = 1 - 0.355$] and $\hat{P}(Z_0 = z_0)$ [$0.499$ for $z_0 = 1$; $0.501$ for $z_0 = 0$]:

\begin{align*}
\hat{E}(Y^{\overline{a}_1=(1,1)})
  &= 153.50 \times 0.645 \times 0.501 \quad (Z_0=0,\;Z_1=0) \\
  &\quad + 152.00 \times 0.355 \times 0.501 \quad (Z_0=0,\;Z_1=1) \\
  &\quad + 155.00 \times 0.446 \times 0.499 \quad (Z_0=1,\;Z_1=0) \\
  &\quad + 153.51 \times 0.554 \times 0.499 \quad (Z_0=1,\;Z_1=1) \\
  &= 49.60 + 27.03 + 34.50 + 42.44 \;=\; 153.57
\end{align*}

Repeating for $\overline{a}_1 = (0,0)$:

\begin{align*}
\hat{E}(Y^{\overline{a}_1=(0,0)})
  &= 150.00 \times 0.732 \times 0.501
   + 148.50 \times 0.268 \times 0.501 \\
  &\quad + 151.50 \times 0.643 \times 0.499
   + 150.00 \times 0.357 \times 0.499 \\
  &= 55.01 + 19.94 + 48.61 + 26.72 \;=\; 150.28
\end{align*}

\noindent giving an estimated average treatment effect of:

\begin{equation}
\hat{\psi} = 153.57 - 150.28 = 3.29
\end{equation}

\noindent Each displayed term equals the product of the (rounded) factors shown; the same computation at full precision also yields $\hat{\psi} = 3.29$. R code to implement this procedure is available in the associated \repolink{}.

\subsection{ICE G-Formula with Two Timepoints}

The ICE g-formula in equation~\eqref{eq:icegform1} proceeds through sequential averaging steps rather than one combined weighted sum. In the case with two time-points, three sequential expectations are taken:

\begin{equation}
E(Y^{\overline{a}_1}) = \underbrace{E \bigg \{ \underbrace{E \bigg [ \underbrace{E \big (Y \mid Z_0, A_0 = a_0, Z_1, A_1 = a_1 \big )}_{\text{first (innermost) expectation}} \, \bigg| \, Z_0, A_0 = a_0 \bigg ]}_{\text{second expectation}} \bigg \}}_{\text{third (outermost) expectation}}
\label{eq:icegform_t2}
\end{equation}

The first (inner) expectation can be thought of as fitting a model to the data, regressing $Y$ against the two treatments and the two time-varying confounders (here, including all of their interactions, i.e., a saturated model), and generating predictions from this model under $A_0 = a_0$ and $A_1 = a_1$ for all individuals in the sample. This procedure represents the innermost expectation.

The second expectation can be thought of as fitting a model that regresses the predictions from the first expectation against $Z_0$ and $A_0$ (again saturated), and generating predictions from this model under $A_0 = a_0$ for all individuals in the sample.

The third (outer) expectation can be thought of as averaging the predictions from the second expectation across all individuals in the sample. Under the causal identification assumptions, this latter expectation can be interpreted as an estimate of $E(Y^{\overline{a}_1})$.

R code to implement this procedure is available in the associated \repolink{}. Running this code with the data in the associated .csv file yields $\hat{E}(Y^{\overline{a}_1=(1,1)}) = 153.57$, $\hat{E}(Y^{\overline{a}_1=(0,0)}) = 150.28$, and thus $\hat{\psi} = 3.29$, identical to the NICE estimate. The exact agreement is expected here: because the models at each stage are saturated (no smoothing is imposed on the 16 strata), both implementations evaluate the same nonparametric quantities, merely organized differently. Had we instead used non-saturated parametric models at each stage (e.g., main-effects-only regressions), the ICE and NICE implementations would generally return slightly different numbers.

\subsection{NICE G-Formula with Multiple Timepoints}

In settings with multiple timepoints, this NICE g-formula expression can be generalized. For instance, following a data generating mechanism such as in Figure \ref{fig:dag_b} generalized to multiple timepoints, we can define a discrete time follow-up period characterized by $t = 0, 1, 2, \ldots T$, where $T$ represents the end of follow-up. We can then write a g-formula for the counterfactual mean:  

\begin{equation}
E(Y^{\overline{a}_T}) = \int \cdots \int E(Y \mid \overline{Z}_T = \overline{z}_T, \overline{A}_T = \overline{a}_T)\prod_{t = 0}^T \,d\mathbb{P}(z_t\mid \overline{z}_{t-1}, \overline{a}_{t-1})
\end{equation}

\noindent where integration occurs at each time point $t$ with respect to the conditional distribution of $Z_t$, and where, by convention, quantities indexed by negative time are empty (so the $t = 0$ factor is the marginal distribution of $Z_0$). In effect, when each $Z_t$ is binary, this equation sums the end-of-follow-up conditional mean of the outcome over all possible confounder histories, weighting each history by the product, across timepoints, of the conditional probabilities of the confounder values given the past. 

\subsection{ICE G-Formula with Multiple Timepoints}

The two-timepoint ICE g-formula in~\eqref{eq:icegform_t2} can be generalized to $T$ timepoints via backward recursion, giving a nested form: 
\begin{equation}
E(Y^{\overline{a}_{T}}) = E\Bigl[ E\bigl[\cdots E\bigl[ E\bigl(Y\mid \overline{Z}_{T},\, \overline{A}_{T}= \overline{a}_{T}\bigr) \mid \overline{Z}_{T-1},\, \overline{A}_{T-1}= \overline{a}_{T-1}\bigr] \cdots \mid Z_0,\, A_0= a_0 \bigr] \Bigr]
\end{equation}
This nested structure naturally leads to a sequential regression estimator. First define:
\begin{equation}
\hat{Q}_{T+1}(\overline{Z}_T) = \hat{E}(Y \mid \overline{Z}_T ,\, \overline{A}_T = \overline{a}_T)
\end{equation}

\noindent where $\hat{Q}_{T+1}(\overline{Z}_T)$ are predictions under a prespecified treatment profile $\overline{a}_T$ from a model that regresses the outcome at time $T$ against everything that comes before it. These predictions can then be iterated backwards for $t = T, T-1, \ldots, 1$:
\begin{equation}
\hat{Q}_t(\overline{Z}_{t-1}) = \hat{E}\bigl(\hat{Q}_{t+1} \mid \overline{Z}_{t-1},\, \overline{A}_{t-1}=\overline{a}_{t-1}\bigr)
\end{equation}

\noindent and recover the predicted counterfactual mean as the final outer expectation:
\begin{equation}
\hat{E}(Y^{\overline{a}_T}) = \hat{Q}_0 = \hat{E}(\hat{Q}_1)
\end{equation}

\noindent where the outermost expectation is over the marginal distribution of $Z_0$. In practice, each step requires regressing $\hat{Q}_{t+1}$ on the observed $(\overline{Z}_{t-1}, \overline{A}_{t-1})$ and predicting under $\overline{A}_{t-1} = \overline{a}_{t-1}$. Note that each regression conditions on (and each prediction is generated under) the \emph{entire} treatment and covariate history up to that point, not merely the most recent values (though in practice, a shorter history is often used). This sequence of $T$ outcome regressions run backward from the end of follow-up is the backbone of the iterative g-computation algorithm and longitudinal targeted maximum likelihood estimation.\cite{Schomaker2019}

\section{Conclusion}

Causal inference methods are based on a foundation of mathematics that can be opaque for applied researchers. This primer has sought to clarify the notation and underlying conceptual foundation behind identifying causal effects using the g-formula, which is obtained by combining the law of iterated expectation with causal identification assumptions.
In addition, the equations representing the law of iterated expectation can be used to estimate causal effects using two versions of the parametric g-computation algorithm, by ``plugging'' predictions from data into the equations resulting from the law of iterated expectation under causal identification assumptions. 


In our examples, the two equivalent representations of this average, NICE and ICE, produce identical estimates by different routes. While this may commonly be the case in time-fixed exposure settings, this will often not be the case in time-varying exposure settings when time-varying confounding is operating. Estimating effects with NICE g-computation requires modeling the distribution of time-varying confounders, where predictions from these models serve as ``weights'' in the law of iterated expectation.\cite{Cole2013, Naimi2021} In contrast, ICE g-computation does not require models for the distribution of time-varying confounders, and is thus robust to misspecification of these models, though at some potential cost in precision.\cite{Wen2021} Intuitively, when the confounder models used by NICE g-computation are correctly specified, they contribute additional information that the classical ICE estimator does not exploit. Simulations suggest that the NICE estimator (and a modified ICE estimator) can be more efficient than the classical ICE estimator described here.\cite{Wen2021,Rudolph2023a}

While ICE g-computation is robust to misspecification of the models for time-varying confounders, both ICE and NICE g-computation require correct parametric models for the outcome as a function of the exposures and confounders. Misspecification of these models can lead to bias, prompting some to use nonparametric (e.g., machine learning) methods instead. However, 
when flexible machine-learning methods are used to estimate nuisance functions, doubly robust procedures such as AIPW or TMLE are often preferred because they can provide valid inference under weaker conditions.\cite{Diaz2020}

Here, we referred to ICE and NICE formulations as different, or alternative, implementations. This is true computationally, in terms of how data are actually used with statistical software code (they are different representations of the same observed data functional), potentially resulting in different numerical values obtained from a g-computation estimator. However, mathematically, ICE and NICE formulations are (nonparametrically) equivalent, meaning they represent the same causal estimand. In our experience, this contrast (on the one hand, they are different, on the other hand, they are equivalent) is often a source of confusion for students encountering these concepts for the first time. Resolving this confusion can serve as an important step in navigating the causal inference literature.

Finally, our focus throughout has been on identification and point estimation. Quantifying sampling uncertainty for g-computation estimators (typically accomplished via the bootstrap) is beyond our scope, and the point estimates reported in our examples are illustrations of the mechanics rather than substantive findings.

Understanding the law of iterated expectation provides a conceptual bridge connecting causal identification to practical estimation. Understanding some of the more technical mathematical concepts that are invoked, as we do in the Appendix, can also serve to clarify important statistical papers in causal inference. Whether implemented in its non-iterative or iterative form (or even augmented with propensity score information to yield doubly robust estimators), the core operation remains a structured, assumption-justified weighted average of conditional outcome means. Familiarity with this operation equips researchers to understand, implement, and critically evaluate g-computation and the doubly robust estimators built upon it.

\section*{Disclosure Statement}

This work was supported by the National Institutes of Health under award numbers K01AI177102 (PNZ), R01AI157758 (PNZ, SRC), and R01HL174652 (AIN). The content is solely the responsibility of the authors and does not necessarily represent the official views of the National Institutes of Health. The authors report no conflicts of interest.

\section*{Data Availability}

The data for Example 1 are reported in full in Table \ref{tab:tamoxifen_data}, from Sato and Matsuyama.\cite{Sato2003} The NHEFS data used in Example 2 are publicly available with the book by Hern\'{a}n and Robins.\cite{Hernan2025} The simulated data used in Example 3, the script that generates them (including the random seed), and R code reproducing all numerical results in this article are available in the associated \repolink{}.

\newpage

\bibliographystyle{epid}
\bibliography{paper_references}

\newpage

\section*{Appendix: Integration, Probability, Statistics, and Measure Theory}

Readers encountering notation such as: 
$$\int g(x)d\mathbb{P}(x)$$

\noindent may naturally wonder how this relates to the ordinary integrals learned in calculus. The answer is that several different notions of integration exist, each designed to generalize the idea of taking weighted averages. While these notions differ mathematically, they share the same underlying goal: accumulating values of a function according to some weighting scheme. 

\begin{table}[ht]
\centering
\caption{Commonly deployed concepts of integration and their typical uses in applied statistics and epidemiology. The choice of integral will often reflect the structure of the random variable and the mathematical tools available.}
\label{tab:int_table}
\begin{tabular}{p{3.2cm}p{3.5cm}p{3.2cm}p{4.5cm}}
\hline
\textbf{Integral Type} & \textbf{Integrates WRT} & \textbf{Intuition} & \textbf{Typical Use} \\
\hline
Finite Sum & $P(X{=}x)$ & Probability mass function & Categorical data, yielding weights for weighted sums\\[6pt]
Riemann & $dx$ & Area under curve & Standard calculus integration \\[6pt]
Riemann--Stieltjes & $dF_X(x)$ & Average using a cumulative distribution function & Classical probability, unifies sum and Riemann integral via CDF increments \\[6pt]
Lebesgue & $d\mu(x)$ & Average with respect to measure & Modern mathematical analysis \\[6pt]
Lebesgue--Stieltjes & $d\nu_{F}(x)$ & Probability distributions & Probability theory \\[6pt]
Probability & $d\mathbb{P}(x)$ & Probability distributions & Most general form used in statistics and causal inference \\[6pt]
\hline
\end{tabular}
\end{table}

\newpage

Commonly deployed concepts of integration, and their typical uses include (Table \ref{tab:int_table}):

\noindent {\bf Riemann Integration}

\noindent The Riemann integral,
$$\int_a^b g(x)dx,$$

\noindent is the integral most readers encounter in introductory calculus. It is constructed by partitioning the domain of the function into many small intervals, approximating the area under the curve using rectangles, and taking a limit as the widths of the rectangles become arbitrarily small. Conceptually, Riemann integration answers the question: How much area lies beneath a curve?

\noindent {\bf Riemann--Stieltjes Integration}

\noindent The Riemann--Stieltjes integral,
$$\int g(x)dF(x),$$

\noindent generalizes the Riemann integral by replacing increments of length $dx$ with increments of a cumulative function $dF(x)$, often a cumulative distribution function (CDF). It is also commonly seen in the causal inference literature (see, for example, Technical Point 1.1 in Hern\'{a}n and Robins\cite{Hernan2025}). Rather than weighting observations according to the width of an interval, the weighting is determined by how much probability accumulates over that interval. When $F$ is a cumulative distribution function, the Riemann--Stieltjes integral accumulates with respect to the probability mass allocated by $F_X$. When $F_X$ is smooth with density $f_X$, we have $dF_X(x) = f_X(x)\,dx$ and the Riemann--Stieltjes integral reduces to the familiar density-weighted integral. When $F_X$ is a step function (as it is for a discrete random variable) the integral becomes a weighted sum. Like the Lebesgue integral, the Riemann--Stieltjes integral provides a single expression that can accommodate both discrete and continuous random variables without modification:
\begin{equation}
E(Y) = \int E(Y\mid X=x)\,dF_X(x),
\end{equation}
which is why this notation appears frequently in the causal inference and statistics literature when authors want one formula to cover discrete and continuous variables alike.

\noindent {\bf Lebesgue Integration}
The Lebesgue integral,
$$\int g(x)d\mu(x),$$

\noindent is the modern foundation of probability theory. Whereas Riemann integration partitions the domain of a function into vertical slices, Lebesgue integration partitions the range of the function into horizontal slices and measures the size of the sets that map into each slice. This construction provides greater mathematical flexibility and supports powerful convergence results that are central to modern statistical theory.

\noindent {\bf Lebesgue--Stieltjes Integration}
The Lebesgue--Stieltjes integral also takes the form
$$\int g(x)dF(x),$$

but is defined using the measure induced by the cumulative distribution function $F$ rather than through the classical Riemann--Stieltjes construction. For most distributions encountered in applied statistics, the numerical value agrees with the corresponding Riemann--Stieltjes integral whenever the latter exists. The distinction becomes important primarily in theoretical work involving convergence, consistency, and asymptotic properties of estimators.

\noindent {\bf Probability Integrals}
\noindent Throughout this paper we use expressions such as
$$\int g(x)d\mathbb{P}(x),$$

\noindent where $\mathbb{P}$ denotes the probability measure induced by a random variable $X$. This notation is best interpreted as taking an average of $g(x)$ with respect to the probability distribution of $X$. When $X$ is discrete,
$$\int g(x)d\mathbb{P}(x) = \sum_x g(x)\,P(X=x).$$

\noindent When $X$ possesses a density $f_X$ with respect to Lebesgue measure,
$$\int g(x)d\mathbb{P}(x) = \int g(x)\,f_X(x)\,dx.$$

\noindent When $X$ contains both discrete and continuous components, the notation remains valid without modification. In particular, the law of iterated expectations and the g-formula can both be expressed compactly using the notation
$$\int g(x)d\mathbb{P}(x),$$

\noindent regardless of whether the covariates are binary, categorical, continuous, or mixtures thereof. Thus, the integral notation $\int E(Y \mid X=x)\,d\mathbb{P}(x)$ used in this primer is deliberately agnostic about the type of variable being averaged over, which is often the notation's primary utility in applied settings. 

However, an important mathematical question involves the kind of mathematical object represented by $d\mathbb{P}(x)$ (the probability distribution governing $X$), and whether this distribution allows us to compute integrals as simple sums, a standard calculus (e.g., Riemann) integral, or whether we require something more general. The answer to this mathematical question depends on the concept of a \emph{dominating measure}, and important concepts in modern analysis and measure theory. 

Every probability distribution $\mathbb{P}$ can be described relative to some reference (dominating) measure, often denoted $\nu$ (pronounced ``nu''). Informally, this reference measure is a rule for assigning ``sizes'' to subsets of the real line. A measure $\nu$ is said to dominate a probability measure $\mathbb{P}$ if $\mathbb{P}$ is absolutely continuous with respect to $\nu$, meaning that every set assigned measure zero by $\nu$ is also assigned probability zero by $\mathbb{P}$. This \emph{absolute continuity} property is written $\mathbb{P} \ll \nu$ (where ``$\ll$'' is read, ``is absolutely continuous with respect to \ldots''). When structured in this measure-theoretic way, we obtain a distribution $\mathbb{P}$ that cannot put a positive probability on any set that $\nu$ considers to have zero size. 
  
These formalisms can be used to derive a central result in measure theory, the Radon-Nikodym theorem, used to guarantee that whenever $\mathbb{P} \ll \nu$ there exists a non-negative function $p = d\mathbb{P}/d\nu$ (the Radon-Nikodym derivative) such that $\int g(x)\,d\mathbb{P}(x) = \int g(x)\,p(x)\,d\nu(x)$. This Radon-Nikodym derivative provides the formal bridge from the dominating measure to the distribution function. The choice of the dominating measure $\nu$ determines which integration framework is being invoked. Common dominating measures correspond to the familiar types of random variable. For a discrete (binary, categorical, or count-valued) random variable, the distribution is dominated by the \emph{counting measure} on the support of $X$, which assigns to any set the number of support points it contains (so each individual support point has size $1$, and any set containing no support points has size $0$). In this case the Radon-Nikodym derivative is the probability mass function $P(X=x)$, and the integral collapses to the familiar weighted sum $\int g(x)\,d\mathbb{P}(x) = \sum_x g(x)\,P(X=x)$. For a continuous random variable, the distribution is typically dominated by the \emph{Lebesgue measure} (a rigorous generalization of ``length'' on the real line), and the Radon-Nikodym derivative is the probability density function $f_X$, so that $\int g(x)\,d\mathbb{P}(x) = \int g(x)\,f_X(x)\,dx$. When $X$ has a mixed distribution (positive probability mass at some points, a density elsewhere), neither measure alone dominates, and a combination of the two is required. 

\end{document}